\documentclass[%
 aps,
 prl,
twocolumn,
superscriptaddress,
frontmatterverbose, 
showpacs,preprintnumbers,
nofootinbib,
 amsmath,
 amssymb,
floatfix,
latexsym,array,enumerate,letter,
]{revtex4-1}

\usepackage[utf8]{inputenc}
\usepackage[english]{babel}
\pdfoutput=1
\allowdisplaybreaks
\usepackage{graphicx,color}
\usepackage[colorlinks=true,citecolor=blue,linkcolor=blue,urlcolor=blue]{hyperref}
\usepackage{comment}
\usepackage{appendix}
\usepackage{footmisc}
\usepackage{bm}
\usepackage{dcolumn}
\usepackage{gensymb}

 \usepackage{float} 
\restylefloat{table}
\usepackage{multirow}
 \usepackage{hyperref}
 \usepackage{array}

\hypersetup{
    colorlinks,
    citecolor=black,
    filecolor=black,
    linkcolor=blue,
    urlcolor=blue
}

\newcounter{infer}[section]
\renewcommand*{\theinfer}{}

\newcolumntype{P}[1]{>{\centering\arraybackslash}p{#1}}
 \newcommand{\be}{\begin{equation}}
\newcommand{\ee}{\end{equation}}
\newcommand{\beq}{\begin{equation}}
\newcommand{\eeq}{\end{equation}}
\newcommand{\bea}{\begin{eqnarray}}
\newcommand{\eea}{\end{eqnarray}}

\usepackage{graphicx,microtype,physics,framed,xcolor}
\usepackage{mdframed,adjustbox} 
\newmdenv[backgroundcolor=gray!15,%
skipabove=0pt,%
skipbelow=5pt,%
leftmargin=0pt,%
rightmargin=0pt,%
innertopmargin=-5pt,%
innerbottommargin=7pt,%
innerleftmargin=2pt,%
innerrightmargin=2pt,%
splittopskip=0pt,%
splitbottomskip=0pt,%
linewidth=0pt,%
nobreak=true]%
{keyeqn}

\begin{document}
\title{Rescaling strange-cluster stars and its implications on gravitational-wave echoes}

\author{Chen Zhang}
\email{iasczhang@ust.hk}
\affiliation{The HKUST Jockey Club Institute for Advanced Study,
The Hong Kong University of Science and Technology, Hong Kong, P.R. China}

\author{Yong Gao}
\email{gaoyong.physics@pku.edu.cn}
\affiliation{Department of Astronomy, School of Physics, Peking University, Beijing 100871, China}
\affiliation{Kavli Institute for Astronomy and Astrophysics, Peking University, Beijing 100871, China}

\author{Cheng-Jun Xia}
\email{cjxia@yzu.edu.cn}
\affiliation{Center for Gravitation and Cosmology, College of Physical Science and Technology, Yangzhou University, Yangzhou 225009, China}

\author{Renxin Xu}
\email{r.x.xu@pku.edu.cn}
\affiliation{Department of Astronomy, School of Physics, Peking University, Beijing 100871, China}
\affiliation{Kavli Institute for Astronomy and Astrophysics, Peking University, Beijing 100871, China}

\begin{abstract}
Solid states of strange-cluster matter called strangeon matter can form strangeon stars that are highly compact. We show that strangeon matter and strangeon stars can be recast into dimensionless forms by a simple reparametrization and rescaling, through which we manage to maximally reduce the number of degrees of freedom. With this dimensionless scheme, we find that strangeon stars are generally compact enough to feature a photon sphere that is essential to foster gravitational-wave (GW) echoes. Rescaling the dimension back, we illustrate its implications on the expanded dimensional parameter space, and calculate the characteristic GW echo frequencies associated with strangeon stars, showing that the minimum echo frequency is $\sim 8$ kHz for empirical parameter space that satisfies the GW170817 constraint, and can reduce to $\mathcal O(100)$ Hertz at the extended limit.
\end{abstract}
\maketitle
\section{Introduction}
Recent gravitational wave (GW) observations from compact binary mergers by LIGO/Virgo collaborations~\cite{LIGOScientific:2016aoc,  LIGOScientific:2017bnn,LIGOScientific:2018mvr,TheLIGOScientific:2017qsa,Abbott:2018wiz,Abbott:2020uma,Abbott:2020khf} have significantly advanced our understanding of black holes and compact stars. These binary merger events have inspired many studies on exotic compact objects (ECOs), which are black hole mimickers that share a similar compactness but lack an event horizon~\cite{Holdom:2016nek,Holdom:2019ouz,Ren:2019afg, Holdom:2022fsm,Mazur:2001fv,Schunck:2003kk, Cardoso:2019rvt}. 
 Most interests on probes of ECOs are focused on the distinctive signatures from gravitational wave echoes in the postmerger signals~\cite{Ignatev:1978ax,Abedi:2018npz, Ferrari:2000sr,Cardoso:2016rao,Cardoso:2016oxy,Cardoso:2017njb,Abedi:2016hgu, Mark:2017dnq,Conklin:2017lwb,Holdom:2020onl, Conklin:2019fcs,Conklin:2019smy, Holdom:2019bdv,Abedi:2020sgg,Dey:2020pth,Dong:2020odp,Wang:2018mlp,Konoplya:2022zym,Bora:2021mba, Bora:2023zhp,Rahman:2021kwb}~\footnote{For probes of ECOs using other methods, see Ref.~\cite{Holdom:2020uhf, Asali:2020wup,Fransen:2020prl,Mukherjee:2022wws}.}, in which a wave that falls inside the gravitational potential barrier (near the photon sphere) travels to a reflecting boundary of the ECO before returning to the barrier after some time delay.

We want to explore the possibility of GW echoes in the context of realistic compact stars, given the detected binary neutron star merger events.  To generate GW echoes, a star object must feature a photon sphere at $R_P=3M$, where $M$ is the object's mass. The minimum radius for compact stars should be above the Buchdahl's limit $R_B=9/4M$~\cite{Buchdahl:1959zz}. Therefore, GW echo signals are possible if $R_B<R<R_P$. This compactness criterion excludes the realistic neutron stars~\cite{Chandrasekhar, Pani:2018flj}. To achieve ultra-compact stellar structure, people commonly assumed \textit{ad hoc} exotic equations of state (EOS)~\cite{Pani:2018flj,Mannarelli:2018pjb, Bora:2020cly,Alho:2021sli,Urbano:2018nrs} or modified gravity~\cite{Volkmer:2021zjx,Bora:2022qwe,Bora:2022dnu}. 

Quark matter, a state comprised of deconfined free-flowing quarks, can possibly exist inside the neutron star core (i.e. hybrid stars~\cite{Alford:2004pf,Alford:2013aca}) or the crust (i.e. inverted hybrid stars~\cite{Zhang:2022pse,Zhang:2023zth}), or constitutes entire star called quark star. Strange quark stars~\cite{Haensel:1986qb,Alcock:1986hz, Zhou:2017pha,Weissenborn:2011qu,Xu:1999bw,Yang:2023haz} composed of strange quark matter (SQM)~\cite{Bodmer:1971we,Witten,Terazawa:1979hq,Farhi:1984qu}  and up-down quark stars~\cite{Zhang:2019mqb,Zhao:2019xqy, Ren:2020tll,Cao:2020zxi,Wang:2021byk,Yuan:2022dxb,Restrepo:2022wqn,Xia:2020byy,Xia:2022tvx,Li:2022vof} composed of up-down quark matter ($ud$QM)~\cite{Holdom:2017gdc}, can be more compact than neutron stars. As Ref.~\cite{Zhang:2021fla} has shown, physically-motivated quark stars can feature GW echoes, but require perturbative QCD corrections~\cite{Fraga:2001id,Fraga:2013qra} and color superconductivity~\cite{Alford:1998mk,Rajagopal:2000ff,Lugones:2002va,Alford:2002kj,Zhang:2020jmb,Zhang:2021iah} effects to be compact enough. It is interesting to explore whether we can have more compact objects from other physical grounds.

Strangeon matter is similar to strange quark matter where both are composed of a nearly equal number of $u,d,s$ quarks~\cite{Xu:2003xe, Lai:2009cn, Lai:2017ney,Miao:2020cqj}.
However, strangeon matter has quarks localized as clusters, in a state more like a solid. Strangeon stars~\cite{Xu:2003xe,Zhou:2004ue,Zhou:2014tba,Lai:2009cn,Lai:2017ney,Miao:2020cqj,Lai:2017mjv,Lai:2017xys,Lai:2018ugk,Lai:2020mlu,Gao:2021uus,Lai:2014hda} composed of strangeon matter have intrinsic stiff EOSs that resulted from the repulsive interaction between strangeons at high densities and the non-relativistic nature of strangeons. The stiff EOS yields large compactness for strangeon stars. They had already been proposed to support massive pulsars ($>2M_\odot$~\cite{Lai:2009cn}) before the announcement of the first massive pulsar PSR J1614-2230~\cite{Demorest:2010bx}. It is then natural to explore whether strangeon stars can feature GW echoes. 

As for the organization of this paper, we first work out a general dimensionless rescaling for the Lennard-Jones model of strangeon matter.  This greatly reduced the number of model parameters from three to one, enabling us to perform a simple but general analysis over the whole parameter space. Then we apply the rescaling scheme to the studies of strangeon stars and GW echoes.

\section{Dimensionless rescaling of Strangeon}
Following previous studies~\cite{Xu:2003xe,Zhou:2004ue,Zhou:2014tba,Lai:2009cn,Lai:2017ney,Miao:2020cqj,Lai:2017mjv,Lai:2017xys,Lai:2018ugk,Lai:2020mlu,Gao:2021uus}, we assume the interaction potential between two strangeons is described by the Lennard-Jones potential~\cite{Jones(1924)}:
\be
U(r)= 4\epsilon \left[ \left(\frac{\sigma}{r}\right)^{12}-\left(\frac{\sigma}{r}\right)^6 \right],
\ee
where $r$ is the distance between two strangeons, and $\sigma$ is the distance when $U(r)=0$. The parameter $\epsilon$ describes the depth of the interaction potential between strangeons. A larger $\epsilon$ will then indicate a larger repulsive force at short range and thus maps to a stiffer EOS.

The mass density $\rho$ and pressure density $p$ of zero-temperature dense matter composed of strangeons derived from Lennard-Jones potential~\cite{Lai:2009cn} reads
\bea\label{eq:energy_density}
    \rho&=&2 \epsilon\left(A_{12} \sigma^{12} n^{5}-A_{6} \sigma^{6} n^{3}\right)
    + nN_{\rm q}m_{\rm q}\,, \\
     p&=&n^{2} \frac{\mathrm{d}(\rho / n)}{\mathrm{d} n}=4 \epsilon\left(2 A_{12}
    \sigma^{12} n^{5}-A_{6} \sigma^{6} n^{3}\right)\,,
\eea
where $A_{12}=6.2$, $A_{6}=8.4$, and $n$ is the number density of strangeons. $N_{\rm q}m_{\rm q}$ is the mass of a strangeon with $N_{\rm q}$ being the
number of quarks in a strangeon and $m_{q}$ being the average constituent quark mass. The contributions from degenerate electrons and vibrations of the
lattice are neglected due to their expected smallness.

At the surface of  strangeon stars, the pressure becomes zero, and we obtain the surface
number density of strangeons as $\big[A_{6}/(2A_{12}\sigma^{6})\big]^{1/2}$. For
convenience, it is transformed into baryon number density, i.e.,
\begin{equation} \label{eq:surface}
    n_{\rm s}=\left(\frac{A_{6}}{2A_{12}}\right)^{1/2}\frac{N_{\rm
    q}}{3\sigma^{3}}\,,
\end{equation}
so that the EOS can be rewritten into a form that depends on parameter set ($\epsilon$, $n_s$,\,$N_q$):
\bea
\rho&=& \frac{1}{9} \epsilon \frac{A_6^2}{A_{12}} \left(\frac{{N_q}^4}{18 n_s^4} n^5 -\frac{ N_q^2}{ n_s^2}  n^3\right)+m_q N_q n, \\
p&=&\frac{2}{9} \epsilon \frac{A_6^2}{A_{12}} \left(\frac{ N_q^4}{9 n_s^4}n^5 -\frac{  N_q^2}{ n_s^2}n^3\right).
\eea
We find that one can further remove the parameters $n_s$ and $N_q$ by doing the following dimensionless rescaling:
\be
\bar{\rho}=\frac{\rho}{m_q \, n_s}, \,\, \bar{p}=\frac{p}{m_q \, n_s},  \, \bar{n}=\frac{N_q \,n }{ n_s},\, \bar{\epsilon}=\frac{\epsilon}{N_q\, m_q}, \,\,
\label{scaling_prho}
\ee
so that
\bea
\bar{\rho}&=& \frac{a}{9} \bar{\epsilon}  \left(\frac{1}{18  } \bar{n}^5 - \bar{n}^3\right)+ \bar{n}, \\
\bar{p}&=&\frac{2 \,a}{9}  \bar{\epsilon}  \left(\frac{1}{9  }\bar{n}^5 -\bar{n}^3\right),
\eea
where $a=A_6^2/A_{12}=8.4^2/6.2\approx11.38$. We thus managed to reduce the parameter degree of freedom from 3 ($n_s$, $\epsilon$, $N_q$) to simply 1 ($\bar{\epsilon}$). 
Besides, we note that the rescaled number density at zero pressure always remain $\bar{n}=3$. Requiring $\bar{\rho}$ to be positive at zero pressure set a theoretical upper bound for $\bar{\epsilon}$: $\bar{\epsilon}_{\rm max}^{\rm theo}=2/a\approx0.1757$. However,  the value of this upper bound is slightly beyond the empirical expectation $\bar{\epsilon}_{\rm max}\sim 120 \text{MeV}/ (3\times 310 \rm MeV)\approx 0.13$. In the following we will adopt the empirical upper bound  $\bar{\epsilon}_{\rm max}^{\rm em}=0.13$ on this physical ground, with additional comments about results from $\bar{\epsilon}_{\rm max}^{\rm theo}$ at proper places.
\section{ReScaling Strangeon Stars}
\begin{figure}[h]
 \centering
\includegraphics[width=8cm]{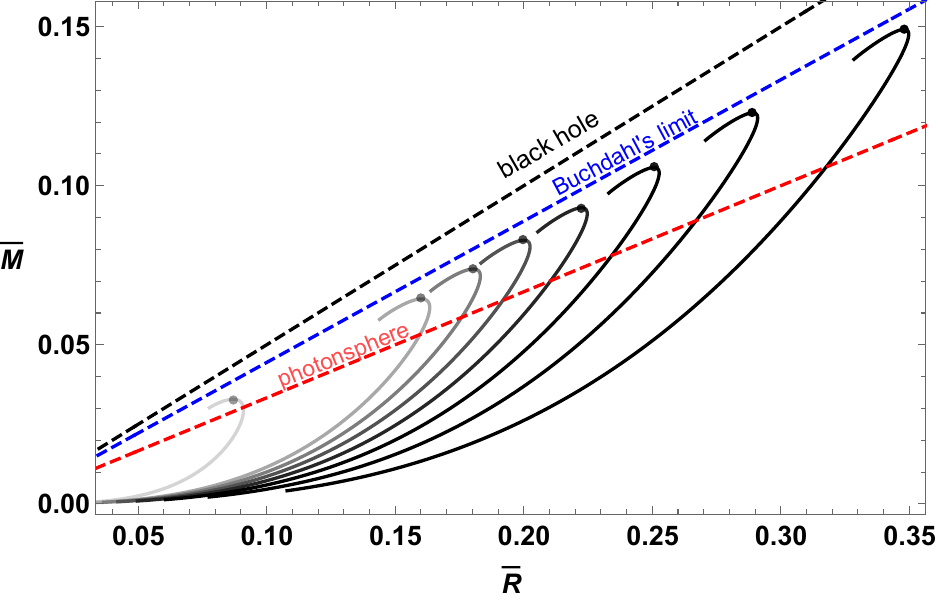}  
\caption{$\bar{M}$-$\bar{R}$ of strangeon stars for various $\bar{\epsilon}$, sampling $0.0001 \sim 0.13$ in equal $\Delta \bar{\epsilon}$ spacing from the lighter black line to the darker black lines, respectively. Solid dots denote the maximum mass configurations. }
   \label{rescaledMR}
\end{figure}

Inspecting the Tolman-Oppenheimer-Volkoff (TOV) equation~\cite{Tolman:1939jz,Oppenheimer:1939ne},
 \bea
 \begin{aligned}
\frac{d m}{d r}& = 4 \pi  \rho r^2\,,\label{eq:dm}\\
\frac{d p}{d r} &= (\rho+p)  \frac{m + 4 \pi p r^3}{2 m r -r^2},\,\\
\end{aligned}
\label{tov}
\eea
we note that the mass and radius can also be rescaled into dimensionless forms in geometric units ($G=c=1$)\footnote{Note that $m_q n_s$, which is in units $\rm MeV/fm^3$ in natural units, is in the dimension of $[L^{-2}]$ in geometric units here.}:
\be
 \bar{m}=m{\sqrt{m_q \, n_s}}, \quad \bar{r}={r}{\sqrt{m_q \, n_s}},
\label{scaling_mr}
\ee
so that the TOV equation can be converted into the dimensionless form (simply replace nonbarred symbols with barred ones). Solving the dimensionless TOV equation, we obtain the results for the rescaled $\bar{M}-\bar{R}$ relation shown in Fig.~\ref{rescaledMR}. 
One can easily recast it into dimensional form by reversing the rescaling relation Eq.~(\ref{scaling_mr}).  
At $\bar{\epsilon}=\bar{\epsilon}_{\rm max}^{\rm em}=0.13$, we have $(\bar{M}_{\rm TOV}, \bar{R}_{\rm TOV})\approx(0.149,\,0.348)$.  Lifting $\bar{\epsilon}$ to $0.175$ that is close to $\bar{\epsilon}_{\rm max}^{\rm theo}$,  we obtain $(\bar{M}_{\rm TOV}, \bar{R}_{\rm TOV})\approx(1.28,\,2.89)$ correspondingly.

\begin{figure}[h]
 \centering
\includegraphics[width=8.5cm]{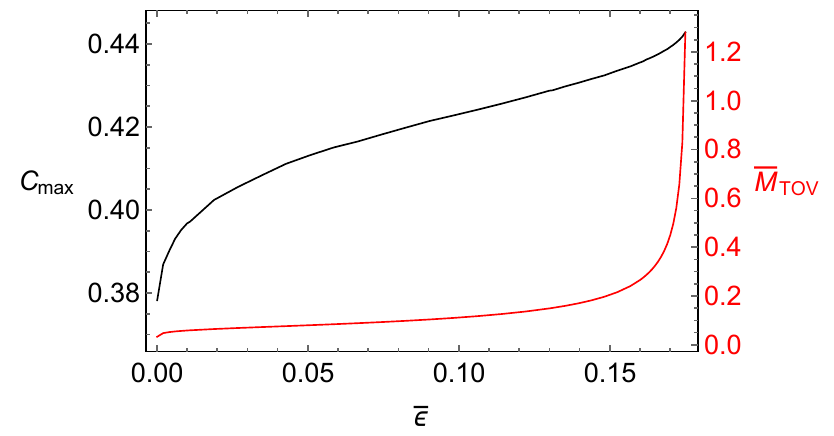}  
\caption{ Maximum compactness $C_{\rm max}=M_{\rm TOV}/R_{\rm TOV}$ (left axis, black) and $\bar{M}_{\rm TOV}$ (right axis, red) of strangeon stars as a function of $\bar{\epsilon}$. }
   \label{CMe}
\end{figure}

The detailed dependences of maximum compactness $M_{\rm TOV}/R_{\rm TOV}$  and the maximum (rescaled) mass $\bar{M}_{\rm TOV}$ on the single parameter $\bar{\epsilon}$ are illustrated more explicitly in  Fig.~\ref{CMe}.\footnote{Note that here we have extended $\bar{\epsilon}$ to $\bar{\epsilon}_{\rm max}^{\rm theo}$ to have a general view.} We find that all $\bar{M}-\bar{R}$ configurations are compact enough to feature a photon sphere while not exceeding the Buchdahl’s limit  for a large range of $\bar{\epsilon}$ variations.\footnote{We examined $\bar{\epsilon}$ as low as $10^{-8}$ order.} Besides, we see clear positive correlations . As $\bar{\epsilon}$ increases to $\bar{\epsilon}_{\rm max}^{\rm theo}$, the maximum mass rapidly grows, while the maximum compactness reaches the Buchdahl's limit.

\begin{figure}[h]
 \centering
\includegraphics[width=8cm]{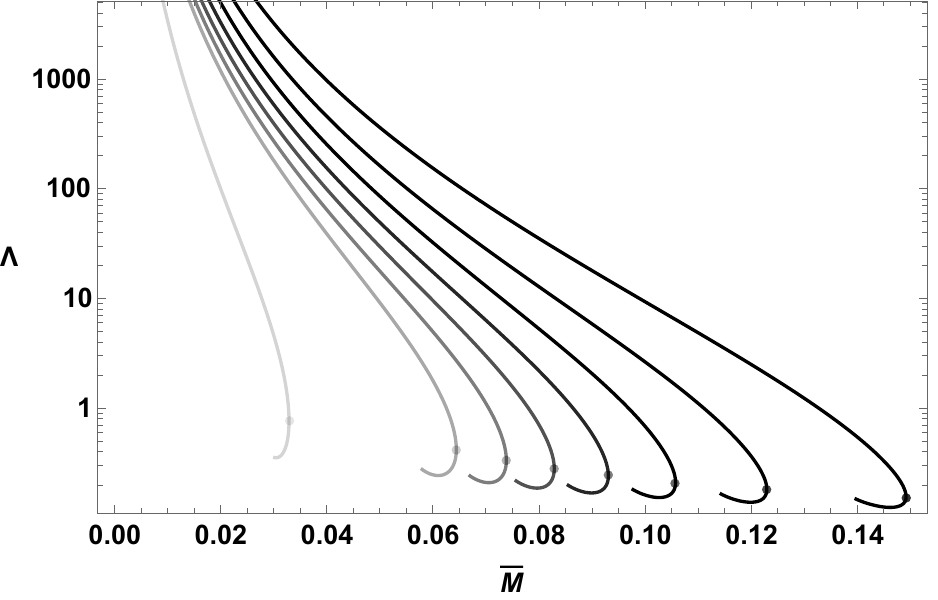}  
\caption{$\Lambda$-$\bar{M}$ of strangeon stars for various $\bar{\epsilon}$. The line-color convention follows that of Fig.~\ref{rescaledMR}. }
   \label{rescaledTidal}
\end{figure}

The rescaled results on tidal deformabilites are shown in Fig.~\ref{rescaledTidal}.  We see that at a given mass, a larger $\bar{\epsilon}$ increases tidal deformability due to a larger radius, as can be observed from Fig.~\ref{rescaledMR}. Besides, we also see that a larger $\bar{\epsilon}$ yields a smaller tidal deformability at the corresponding maximum mass point due to the associated larger compactness.

\section{GW Echoes from Strangeon Stars}
\begin{figure}[h]
 \centering
\includegraphics[width=8cm]{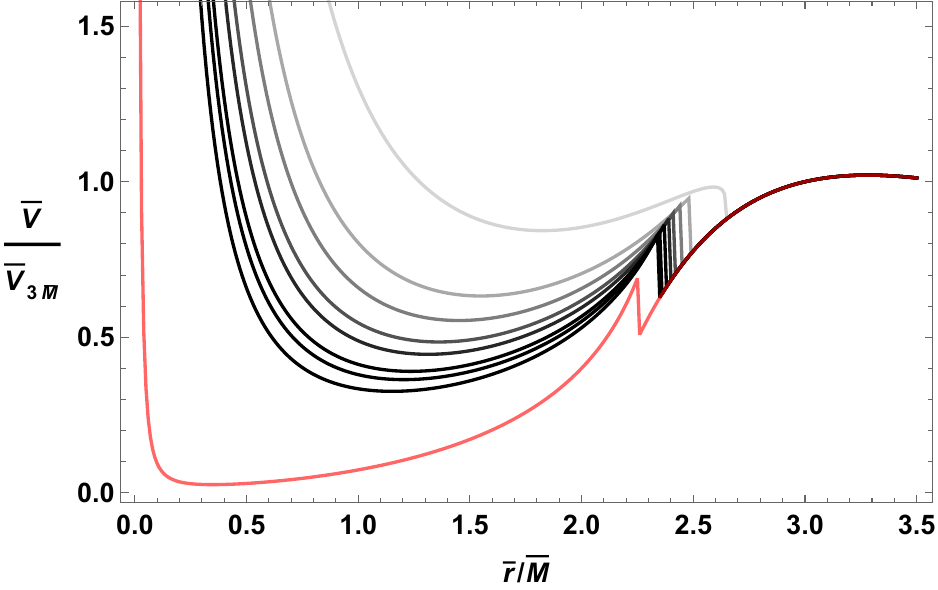}  
\caption{Radial profiles of effective potentials for axial gravitational perturbations of the $ l=s=2$ mode in strangeon-star background  at $M_{TOV}$ points for various $\bar{\epsilon}$. The color convention of black lines follows that of Fig.~\ref{rescaledMR}. The red line denotes the $\bar{\epsilon}=0.175\approx\bar{\epsilon}_{\rm max}^{\rm theo}$ limit. }
   \label{rescaled_V}
\end{figure}

 The effective potential of axial gravitational  perturbations $\Psi_{s,l}$ in curved background has the general form~\cite{Conklin:2017lwb,Urbano:2018nrs}:
\bea\label{eq:potential}
V(r)&=&B(r) \Big\{ \frac{l(l+1)}{r^2}+\frac{1-s^2}{2rA(r)}\left(\frac{B^\prime(r)}{B(r)}-\frac{A^\prime(r)}{A(r)}\right) \nonumber\\
&+& 8\pi (p(r)-\rho(r))\delta_{s,2} \Big\},
\eea
 where  the azimuthal quantum number $ l \geq s$ with $s = 0,\pm 1,\pm2$ for scalar, vector and tensor modes, respectively. And
 \be
 B(r)=e^{2\Phi(r)}\,,\, A(r)=\frac{1}{1-2m(r)/r}
 \ee
as the metric factors of curved line element describing spherical symmetric spacetime: $ds^2 = -B(r)dt^2 + A(r)dr^2 + r^2d\theta^2 + r^2 \sin^2 \theta d\phi^2$. $\Phi(r)$ is solved via
\be
\frac{ d \Phi}{dr}  =-\frac{1}{\rho+p} \frac{d p}{d r},
\label{Phi}
\ee
together with the TOV equation Eq.~(\ref{tov}).  Apparently, we can apply rescaling relation Eq.~(\ref{scaling_mr}) and Eq.~(\ref{scaling_prho}), and perform a further rescaling $\bar{V}=V/(m_q\,n_s)$  to convert the whole program into a dimensionless form as other barred quantaties.   We obtain the effective potential of the lowest axial gravitational perturbation mode ($l=s=2$) in strangeon star background shown in Fig.~\ref{rescaled_V},\footnote{Note that we normalized $\bar{V}(\bar{r})$ respect to $\bar{V}(\bar{r}=3\bar{M})$, and $\bar{r}$ respect to $\bar{M}$, where the rescaling factors cancel and thus would yield same result for the dimensional version.} which abruptly changes at the star surface, diverges towards star center with an outside peak around $3\bar{M}$, forming a trapping cavity for gravitational waves.  We  see clearly the trend how the trapping cavity develops and evolves  as  the parameter $\bar{\epsilon}$  increases. 

The characteristic echo time is the light time from the star center to the photon sphere~\cite{Cardoso:2017njb,Cardoso:2016rao,Cardoso:2016oxy},
\be
\tau_\text{echo} = \int_0^{3M} \hspace{-.5cm}\frac{dr}{\sqrt{e^{2\Phi (r)}\left(1- \frac{2 m(r)}{r}\right)}}\,,
\label{tau_echo}
\ee

\begin{figure}[h]
 \centering
\includegraphics[width=8cm]{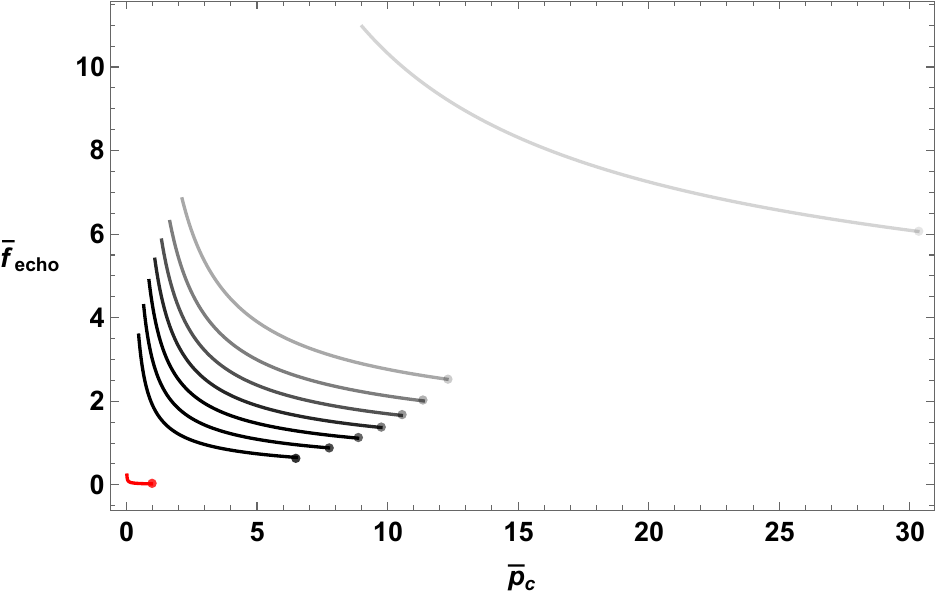}  
\caption{Rescaled echo frequencies $\bar{f}_{\rm echo}$ as functions of center pressure $\bar{p}_c$ of strangeon stars for various $\bar{\epsilon}$. The color convention of black lines follows that of Fig.~\ref{rescaledMR}. The red line denotes the $\bar{\epsilon}=0.175\approx\bar{\epsilon}_{\rm max}^{\rm theo}$ limit. }
   \label{rescaled_fpc}
\end{figure}

We can also do the dimensionless rescaling
\be
\quad \bar{\tau}_\text{echo} ={\tau}_\text{echo} {\sqrt{m_q \, n_s}}, \,
\ee
such that Eq.~(\ref{tau_echo}) can also be calculated in a dimensionless approach. After obtaining the echo time, we directly get the GW echo frequency from the relation~\cite{Cardoso:2017njb,Cardoso:2016rao,Cardoso:2016oxy} 
\be
f_\text{echo}=\frac{\pi}{\tau_\text{echo}}, 
\label{ftau}
\ee 
and similarly, we can rescale it into the dimensionless form $\bar{f}_\text{echo} $ via the relation 
\be
\bar{f}_\text{echo} =\frac{{f}_\text{echo}}{\sqrt{m_q \, n_s}}.
\label{scaling_f}
\ee
In Fig.~\ref{rescaled_fpc}, we show the results of rescaled GW echo frequencies $\bar{f}_\text{echo}$ versus the rescaled center pressure $\bar{p}_c$ for the stellar configurations of Fig.~\ref{rescaledMR} that can generate echoes. Note that each curve's left and right ends are truncated at the point where $\bar{R}=3\bar{M}$, and at the point of maximum mass, respectively. 

As $\bar{\epsilon}$ and the central pressure $\bar{p}_c$ increases,  $\bar{f}_\text{echo}$  decreases (due to the increasing compactness) with a lower bound $\bar{f}_\text{echo}^{\,\rm min}(\bar{\epsilon})$ set at the $\bar{p}_c$ of the maximum mass point. At $\bar{\epsilon}=\bar{\epsilon}_{\rm max}^{\rm em}=0.13$, we have the minimum echo frequency $\bar{f}_{\rm echo}^{\rm min}=0.655$, translating to  
\be
f_{\rm echo}^{\rm min}\approx 2 \left(\frac{n_s}{0.24\rm\, fm^{-3}} \right)^{1/2} \rm \, kHz  \quad \text{for } \bar{\epsilon}=\bar{\epsilon}_{\rm max}^{\rm em}.
\ee
Lifting $\bar{\epsilon}$ to $0.175$ that is close to $\bar{\epsilon}_{\rm max}^{\rm theo}$,  we obtain $\bar{f}_{\rm echo}^{\rm min}=0.023$, mapping to
\be
f_{\rm echo}^{\rm min}\approx 67 \left(\frac{n_s}{0.24\rm\, fm^{-3}} \right)^{1/2} \rm \, Hz  \quad \text{for } \bar{\epsilon}\approx\bar{\epsilon}_{\rm max}^{\rm theo}.
\ee
Interestingly, we see that in this extreme limit, the minimum echo frequencies lie well within the sensitivity range of LIGO~\cite{LIGOScientific:2016aoc,Urbano:2018nrs}.

\section{Dimensional Parameter Space}

\begin{figure}[htb]
 \centering
\includegraphics[width=8.2cm]{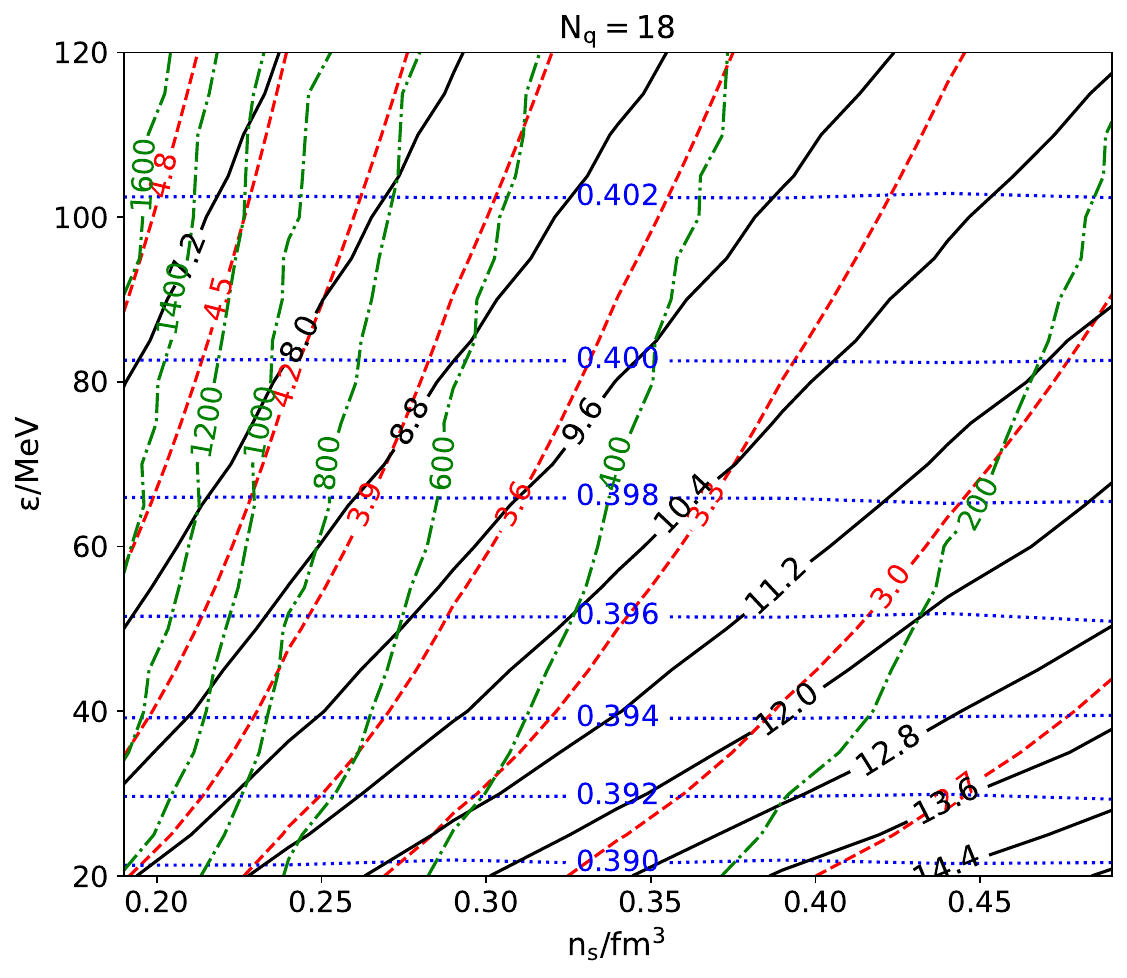}  
\includegraphics[width=8.2cm]{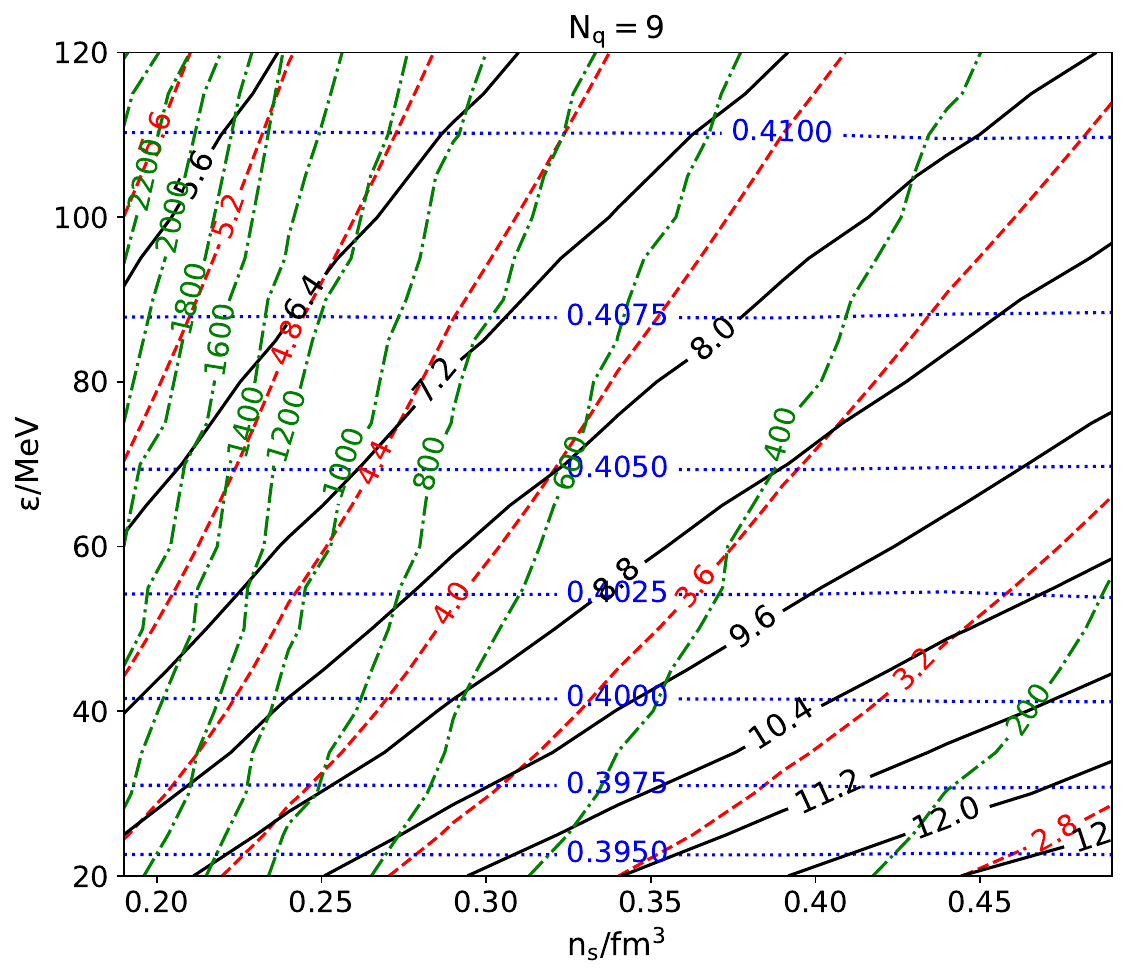}  
\caption{Physical parameter space for (top) $N_q=18$ and (bottom) $N_q=9$. Black lines denote $f_{\rm echo}/\text{kHz}$, with red lines denoting the maximum masses $M_{\rm TOV}/ M_{\odot}$, green lines for the tidal deformabilities $\Lambda$ at $1.4\, M_{\odot}$ and blue dotted lines for the maximal compactness $M_{\rm TOV}/R_{\rm TOV}$.}
   \label{paraspace}
\end{figure}

In Fig.~\ref{paraspace}, we show the derived quantites in dimensional forms ($f_{\rm echo}$, $M_{\rm TOV}$) and ($\Lambda$, $C$) in dimensional parameter space of ($\epsilon$, $n_s$) by rescaling back previous simple dimensionless results using relations  Eq.~(\ref{scaling_prho}) and Eq.~(\ref{scaling_mr}).

Figure~\ref{paraspace} manifests the apparent scaling behaviour:
\begin{itemize}
\item Decreasing $\epsilon$ for given $N_q$ or increasing $N_q$ for given $\epsilon$ is equivalent in terms of $\bar{\epsilon}$. 
\item Dimensionless quantities like the compactness $C=M/R$ should be independent of $n_s$. This explains why the blue dotted lines are flat.
\item Dimensional quantities follows the scaling relation dictated by Eq. ~(\ref{scaling_prho}) and Eq.~(\ref{scaling_mr}). For example, the maximum mass $M_{\rm TOV}$ (and corresponding radius $R_{\rm TOV}$) scale as $\sqrt{n_s}$.
\end{itemize}

From Fig.~\ref{paraspace}, we see explicitly that the minimum echo frequency is $\sim 8$ kHz for parameter space of large $\epsilon$ and small $n_s$ while satisfies the GW170817 tidal deformability constraint $\Lambda({1.4\,M_\odot})\lesssim 800$, and can be reduced to $5$ kHz if GW170817 constraint is dropped, i.e., if assuming the detected stars in binary merger not strangeon stars.

\section{Summary}
 We worked out a first rescaling scheme that enables us to maximally reduce the number of free parameters into a single parameter $\bar{\epsilon}$ for strangeon matter. Utilizing this scheme, we demonstrated that strangeon stars composed of strangeon matter generally have very large compactness with large $\bar{\epsilon}$ in most of its parameter space. We showed that all strangeon stars can meet the compactness condition for generating GW echoes, i.e., they feature a photon sphere within Buchdahl's limit.  
 The minimum echo frequencies are of a few kilohertz for the empirical range of $\bar{\epsilon}$, and can reduce to $O(100)$ Hertz if $\bar{\epsilon}$ is extended to its allowed limit. We explicitly constructed the corresponding dimensional parameter space of $\epsilon$ and $n_s$ with variations of $N_q$ in their empirical range, and demonstrated that $f_\text{echo}^{\rm min}\approx 8\, \rm kHz$ for realisitic parameter space that satisfy astrophysical constraints like GW170817, and can reduce to $5$ kHz if the latter constraint is dropped.

It is generally expected that including the star-rotation effect can slightly reduce the echo frequencies~\cite{Pani:2018flj,Urbano:2018nrs}. For strangeon stars, we expect rotation would yield a similar reduction of $f_{\rm echo}$, potentially reducing frequencies to what LIGO can detect, considering our obtained  $f_\text{echo}^{\rm min}\approx 5\sim 8 \rm\, kHz$ for the realistic non-rotating case is not very far from its detection limit.  Besides, strangeon stars may have hybrid structures, likely a strangeon matter crust and a strange quark matter core resulting from a first-order phase transition at high densities~\cite{Zhang:2023szb}. Such a new type of hybrid stars may be also compact enough to signal GW echoes. We leave these interesting possibilities for future studies.

\begin{acknowledgments}
\noindent\textbf{Acknowledgments. }  
C. Zhang greatly thank Prof. Renxin Xu for the visit invitation to Peking University and is very grateful for the hospitality during the visit.  C. Zhang is supported by the Institute for Advanced Study at The Hong Kong
University of Science and Technology.  C.~J Xia is supported by National Natural Science Foundation of China (Grant No. 12275234). Y. Gao and C. J Xia are supported by National SKA Program of China (Grant No. 2020SKA0120300). R.X Xu is supported by the National SKA Program of China (2020SKA0120100).

\end{acknowledgments}

\end{document}